\documentclass[]{IEEETran}
\usepackage{cite}
\usepackage{amsmath,amssymb,amsfonts}
\usepackage{algorithmic}
\usepackage{graphicx}
\usepackage{textcomp}
\usepackage{xcolor}
\usepackage{subfig}
\usepackage{caption}
\usepackage{bm}
\usepackage{xspace}

\newcommand{\bmath}[1]{\ensuremath{\bm{#1}}\xspace}

\newcommand{\tht}{\bmath{\theta}}

\newcommand{\kap}{\bmath{\kappa}}
\newcommand{\ro}{\bmath{\rho}}

\newcommand{\muv}{\bmath{\mu}}

\newcommand{\U}{\bmath{U}}

\newcommand{\beq}{\begin{equation}}
\newcommand{\eeq}{\end{equation}}
\newcommand{\bea}{\begin{eqnarray}}
\newcommand{\eea}{\end{eqnarray}}
\newcommand{\ba}{\left(\!\!\begin{array}}
\newcommand{\ea}{\end{array}\!\!\right)}
\newcommand{\bc}{\begin{center}}
\newcommand{\ec}{\end{center}}

\def\BibTeX{{\rm B\kern-.05em{\sc i\kern-.025em b}\kern-.08em
    T\kern-.1667em\lower.7ex\hbox{E}\kern-.125emX}}

\begin{document}
	\title{{Incorporating Tissue Composition Information in Total-Body PET Metabolic Quantification of Bone Marrow through Dual-Energy CT}}
	\author{Siqi Li, Benjamin A. Spencer, Yiran Wang, Yasser G. Abdelhafez, Heather Hunt, J. Anthony Seibert, \\ Simon R. Cherry, Ramsey D. Badawi, Lorenzo Nardo, and Guobao Wang
		
	\thanks{This work is supported by NIH R01EB036562. The study was also supported in part by NIH R21 EB027346 and In Vivo Translational Imaging Shared Resources with funds from NCI P30CA093373. The University of California, Davis, has a research agreement and revenue-sharing agreement with United Imaging Healthcare. No other potential conflict of interest relevant to this article was reported.
		}
		\thanks{This work involved human subjects or animals in its research. Approval of all ethical and experimental procedures and protocols was granted by Institutional Review Board (IRB) at the University of California, Davis and written informed consent was obtained for all study participants.
		}
		\thanks{Siqi Li, Benjamin A. Spencer, Yiran Wang, Yasser G. Abdelhafez, Heather Hunt, J. Anthony Seibert, Lorenzo Nardo, and Guobao Wang are with the Department of Radiology, University of California Davis Medical Center, Sacramento, CA 95817 USA. (e-mail: sqlli@ucdavis.edu, benspencer@ucdavis.edu, yrdwang@ucdavis.edu, yabdelhafez@ucdavis.edu, hlhunt@ucdavis.edu, jaseibert@ucdavis.edu, lnardo@ucdavis.edu, gbwang@ucdavis.edu).}
		\thanks{Simon R. Cherry and Ramsey D. Badawi are with the Department of Radiology and Department of Biomedical Engineering, University of California Davis, Davis, CA 95616 USA (e-mail: srcherry@ucdavis.edu, rdbadawi@ucdavis.edu).}}
	\maketitle

\begin{abstract}
Bone marrow (BM) metabolic quantification with $^{18}$F-fluorodeoxyglucose (FDG) positron emission tomography (PET) is of broad clinical significance for accurate assessment of BM at staging and follow-up, especially when immunotherapy is involved.  However, current methods of quantifying BM may be inaccurate because the volume defined to measure bone marrow may also consist of a fraction of trabecular bone in which $^{18}$F-FDG activity is negligible, resulting in a potential underestimation of true BM uptake. In this study, we demonstrate this bone-led tissue composition effect and propose a bone fraction correction (BFC) method using X-ray dual-energy computed tomography (DECT) material decomposition. This study included ten scans from five cancer patients who underwent baseline and follow-up dynamic $^{18}$F-FDG PET and DECT scans using the uEXPLORER total-body PET/CT system. The voxel-wise bone volume fraction was estimated from DECT and then incorporated into the PET measurement formulas for BFC. The standardized uptake value (SUV), $^{18}$F-FDG delivery rate $K_1$, and net influx rate $K_i$ values in BM regions were estimated with and without BFC and compared using the statistical analysis. The results first demonstrated the feasibility of performing voxel-wise material decomposition using DECT for metabolic BM imaging. With BFC, the SUV, $K_1$, and $K_i$ values significantly increased by an average of 13.28\% in BM regions compared to those without BFC (all P$<$0.0001), indicating the impact of BFC for BM quantification. Parametric imaging with BFC further confirmed regional analysis. Our study using DECT suggests current SUV and kinetic quantification of BM are likely underestimated in PET due to the presence of a significant bone volume fraction. Incorporating tissue composition information through BFC may improve BM metabolic quantification.

	\end{abstract}
		\vspace{35pt}
	\section{Introduction}
	\IEEEPARstart{T}{he} evaluation of bone marrow quantification using $^{18}$F-fluorodeoxyglucose (FDG) positron emission tomography (PET) imaging is of great clinical importance, particularly for diagnosing, staging, and managing various malignancies. For example: (i) In hematological disorders, such as non-Hodgkin lymphoma, leukemia, multiple myeloma and other cancer types, identifying bone marrow involvement is crucial for accurate cancer staging and assessing tumor prognosis \cite{Adams2014, Y2019, Khan2013}. (ii) In the context of anti-cancer immunotherapy, the metabolic uptake observed in bone marrow has emerged as a prognostic biomarker, offering valuable insights for the assessment of therapeutic responses\cite{Nakamoto2021, Berthet2013}. Beyond oncology, metabolic PET quantification of bone marrow may also be useful in evaluating infectious diseases \cite{Pijl2020}, systemic inflammation \cite{Patel2021}, and autoimmune conditions \cite{Zhang2022}, where altered marrow activity may reflect underlying immune activation.
	
\begin{figure*}[t]
		\footnotesize
		\centering
		{\includegraphics[trim=0cm 0cm 0cm 0cm, clip,width=5in]{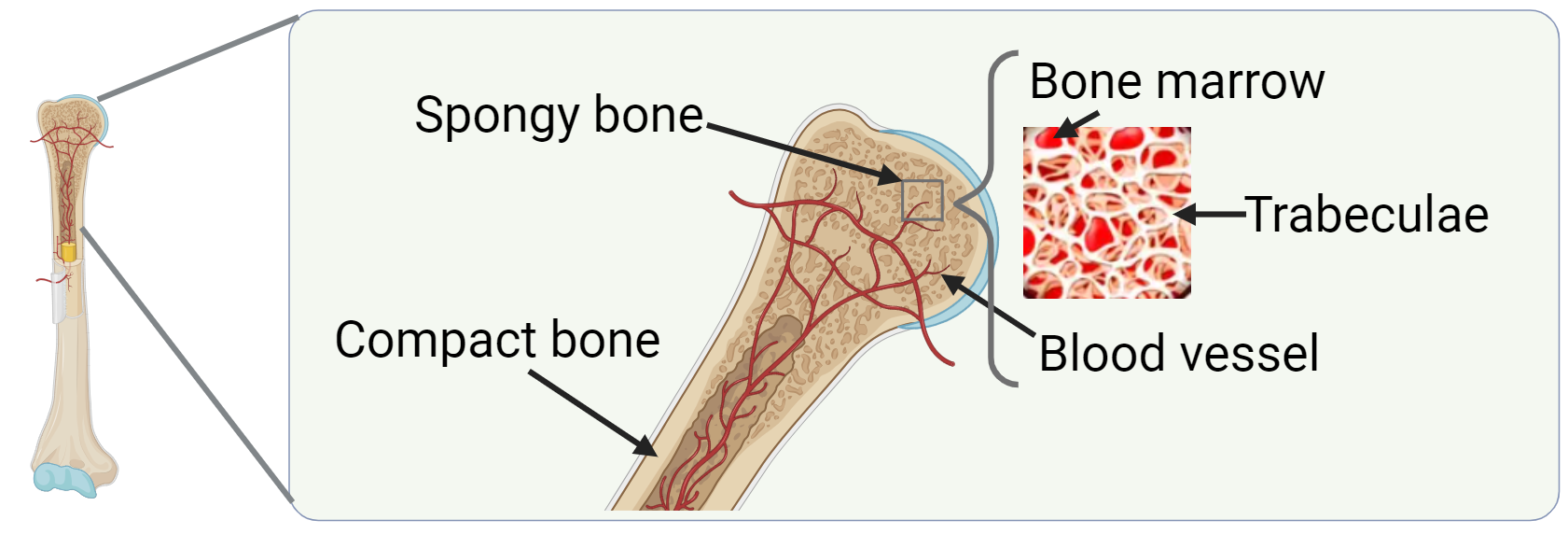}}
		\caption{A graphical illustration of bone anatomy structure with spongy bone. A unit volume of “bone marrow” defined by PET may include actual bone marrow, trabeculae bone, and blood vessels.}
		\label{OBM}
	\end{figure*}

	However, the current methods used with PET to assess bone marrow may be inaccurate. This inaccuracy arises from the existing method used to calculate radiotracer uptake within a unit volume, which assumes that this volume contains only actual bone marrow. Owing to the inherently limited spatial resolution of PET, a single voxel in the bone marrow region will contain not only the true bone marrow itself but also surrounding tissue structures, such as trabecular bone and blood vessels, as illustrated in Fig. \ref{OBM}. It is important to note that the uptake of $^{18}$F-FDG  in the trabecular bone can be negligible. Consequently, the actual metabolic uptake within the true bone marrow may be significantly underestimated when ignoring the impact of the bone volume fraction. This underestimation may affect the accuracy of bone marrow quantification in scenarios where changes in bone density may occur, such as aging, bone-related diseases, cancer treatment, and other conditions.
	
	X-ray dual-energy computed tomography (DECT) imaging \cite{McCollough2015,McCollough2020} can provide bone fraction information which may complement PET imaging of bone marrow. It uses two distinct X-ray energies to obtain energy-dependent tissue attenuation information, enabling quantitative multi-material decomposition. One general application of DECT is to generate a virtual non-calcium image, which improves the visualization of soft-tissue regions by removing bone-similar components from the CT image \cite{Murphy2016,Virarkar2022}. Existing work combining DECT with PET mainly focus on improving the attenuation correction for quantitative PET imaging \cite{Noh2009,Xia2013} or providing other imaging metrics (e.g., tissue composition) beyond PET measurements for tumor detection and characterization  \cite{Gaudreault2023,Nagano2022}. However, the potential of DECT for correcting bone marrow quantitation in PET has not been demonstrated.

	In this work, we aim to investigate this bone-led tissue composition effect using DECT material decomposition and evaluate its impact on metabolic PET quantification of bone marrow with the total-body uEXPLORER PET/CT scanner \cite{Cherry2017,Badawi2019,Spencer2021}. In addition to the standardized uptake value (SUV) measured by static PET, we also investigate the effect of bone fraction on kinetic quantification. Dynamic PET imaging with tracer kinetic modeling enables multiparametric imaging \cite{Wang2020a} and has the potential to comprehensively evaluate the quantitative characterization of bone marrow, such as in multiple myeloma \cite{Sachpekidis2015} or COVID-19 studies \cite{Wang2023, Omidvari2023}. The advent of total-body PET scanners provides a unique opportunity for a robust kinetic analysis of bone marrow while also allowing us to study the effect of bone fraction on the quantification of different bone marrow regions throughout the whole body \cite{Wang2021}. This work will primarily focus on the conceptual demonstration and technical development of bone fraction correction (BFC) method, showing how BFC impacts the static and kinetic quantification of bone marrow. 
	
	A portion of this work was presented at the 2022 Society of Nuclear Medicine and Molecular Imaging (SNMMI) conference \cite{Li2022}. Here we detailed the theoretical methodology for correcting metabolic quantification of bone marrow quantification in PET using DECT and evaluated the impact of bone volume fraction on more patient studies. This article is organized as follows. Section II describes the theoretical effect of bone marrow on both static and dynamic PET imaging. Section III introduces the total-body PET/DECT datasets and our evaluation strategies. We then present the results in Section IV and follow with a comprehensive discussion of the findings and limitations of this study in Section V. Finally, conclusions are drawn in Section VI.

	\section{Proposed BFC Method}
	\subsection{Theoretical Effect of Bone Fraction on Static FDG-PET Quantification}
Traditionally, the calculation of SUV in bone marrow assumes that each unit volume is entirely true bone marrow, i.e.:
\beq
\text{SUV}_{\text{noBFC}} = \frac{\text{SUV}_{\text{meas}}}{1},
\eeq
where $\text{SUV}_{\text{meas}}$ is the measured $^{18}$F-FDG SUV in PET.

However, as shown in Fig. \ref{OBM}, it is important to note that a unit volume of “bone marrow” in PET images can also include the trabecular bone, where the $^{18}$F-FDG uptake can be negligible, due to the limited spatial resolution of PET. Thus, a more accurate bone marrow SUV should account for the fractional volume of trabecular bone $v_{\rm{bone}}$,
\beq
\text{SUV}_{\text{BFC}} = \frac{\text{SUV}_{\text{meas}}}{1 - v_{\rm{bone}}},
\label{SUV_BFC}
\eeq
where  $\text{SUV}_{\text{BFC}}$ is the corrected SUV value in the actual bone marrow volume (i.e., $1 - v_{\rm{bone}}$). If $v_{\rm{bone}}$ is large, this oversight may cause a significant underestimation of actual bone marrow SUV.

	\begin{figure*}[t]
	\vspace{-0pt}
	\footnotesize
	\centering
	{\includegraphics[trim=0cm 0cm 0cm 0cm, clip,width=5in]{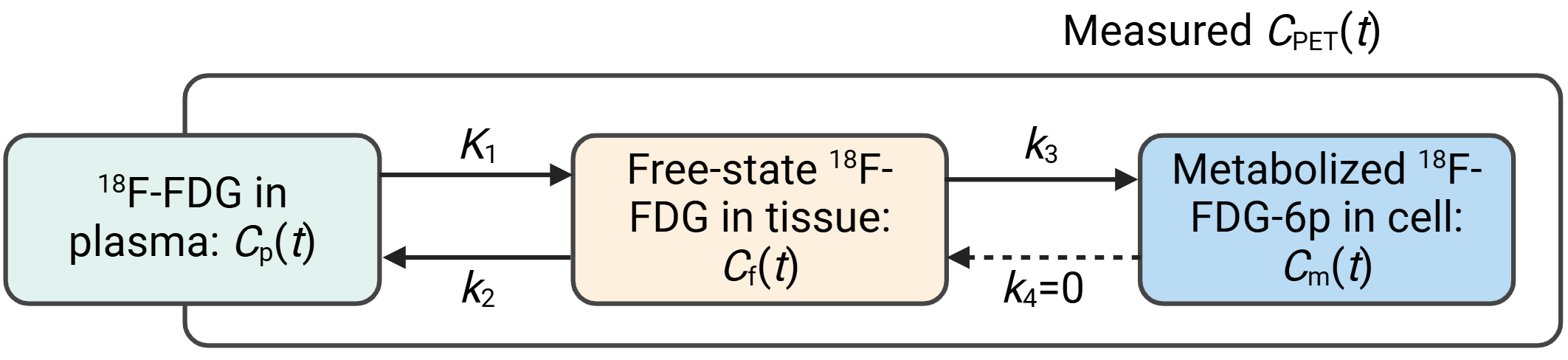}}
	\caption{The 2-tissue irreversible compartmental model for describing $^{18}$F-FDG kinetics of bone marrow.}
	\label{fig:2}
	\vspace{-0pt}
\end{figure*}
\subsection{Theoretical Effect of Bone Fraction on Dynamic FDG-PET Kinetic Quantification}
\subsubsection{Model of tissue compositions for bone marrow}
In addition to the true bone marrow and trabecular bone, a fraction of a unit volume seen as “bone marrow” on a PET image (denoted as $v_{\rm{PET}}$) is the associated vascular volume, as illustrated in Fig. \ref{OBM}. We thus have the following formula for $v_{\rm{PET}}$,
	\beq
	v_{\rm{PET}} = v_{\rm{BM}} + v_{\rm{bone}} + v_{\rm{blood}} = 100\%,
	\label{volume}
	\eeq
	where $v_{\rm{BM}}$ and $v_{\rm{blood}}$ represent the fractional volumes of the bone marrow and whole blood, respectively. The measured $^{18}$F-FDG concentration $C_{\rm{PET}}$ can thus be represented as a volume-weighted summation,
	\beq
	C_{\rm{PET}} =v_{\rm{BM}}C_{\rm{BM}}+v_{\rm{bone}}C_{\rm{bone}}+v_{\rm{blood}}C_{\rm{wb}},
	\eeq
	where  $C_{\rm{BM}}$, $C_{\rm{bone}}$, and $C_{\rm{wb}}$ represent the $^{18}$F-FDG uptake in the true bone marrow, trabecular bone, and whole blood, respectively. Since $v_{\rm{BM}}$ is equal to $1-v_{\rm{bone}}-v_{\rm{blood}}$ (Eq. \ref{volume}) and $C_{\rm{bone}}$ can be approximated as 0, $C_{\rm{PET}}$ is further reduced to
	\beq
	C_{\rm{PET}} =(1-{v_{\rm{bone}}}-v_{\rm{blood}}){C_{\mathrm{BM}}}+v_{\rm{blood}}C_{\rm{wb}}.
	\label{C_PET}
	\eeq

\subsubsection{Compartmental models with BFC}
Following the commonly used 2-tissue irreversible compartmental model shown in Fig. \ref{fig:2} \cite{MORRIS2004} and Eq. \ref{C_PET}, the bone fraction-corrected $^{18}$F-FDG activity at time $t$, $C_{\rm{PET}}(t)$, can be modeled as follows,
\beq
C_{\rm{PET}}(t) = (1 - v_{\rm{bone}} - v_{\rm{blood}})\underbrace{(C_{\rm{f}(t)} + C_{\rm{m}(t)})}_{C_{\mathrm{BM}}(t)} + v_{\rm{blood}} C_{\rm{wb}}(t),
\label{C_PETt}
\eeq
where $C_{\rm{f}}(t)$ and $C_{\rm{m}}(t)$ represent the concentrations of tissue free-state $^{18}$F-FDG and tissue-metabolized $^{18}$F-FDG at time $t$, respectively. $C_{\rm{wb}}(t)$ here is approximately equal to the plasma input function $C_{\rm{p}}(t)$. We then have the following corresponding ordinary differential equations to describe the system states,
	\beq
\
\frac{d}{dt} 
\begin{bmatrix}
	C_{\rm{f}}(t) \\
	C_{\rm{m}}(t)
\end{bmatrix}
=
\begin{bmatrix}
	-(k_2+k_3) & 0 \\
	k_3 & 0
\end{bmatrix}
\begin{pmatrix}
	C_{\rm{f}}(t) \\
	C_{\rm{m}}(t)
\end{pmatrix}
+
\begin{bmatrix}
	K_1 \\
	0
\end{bmatrix}
C_{\rm{p}}(t),
\,
\eeq
where $K_1$ (mL/min/cm$^3$) represents the $^{18}$F-FDG plasma-to-tissue delivery rate, and $k_2$ (min$^{-1}$) is the tissue-to-plasma delivery rate. $k_3$ (min$^{-1}$) is the $^{18}$F-FDG phosphorylation rate, and the irreversible model assumes that the $^{18}$F-FDG dephosphorylation rate can be negligible, i.e. $k_4=0$. With the Laplace transform, Eq. \ref{C_PETt} can be further represented as
	\beq
C_{\rm{PET}}(t) = (1 - v_{\rm{bone}} - v_{\rm{blood}}) H(t;\kap)\otimes C_{\rm{p}}(t) + v_{\rm{blood}} C_{\rm{wb}}(t).
\eeq
In this equation, $H(t;\kap) = \frac{K_1}{k_2+k_3}(k_3+k_{2}e^{-(k_2+k_3)t})$, which represents the impulse response function of the 2-tissue irreversible model with $\kap=[K_1, k_2, k_3]^{\rm{T}}$.
\subsubsection{Time delay correction and model fitting}
It is worth noting that the time delay correction is important for total-body kinetic modeling \cite{Wang2021,Wang2022a}, we thus model the time delay effect into the input function as
	\begin{equation}
	\begin{aligned}
		C_{\rm{PET}}(t) = & (1 - v_{\rm{bone}} - v_{\rm{blood}})H(t;\kap) \otimes C_{\rm{p}}(t-t_d) \\ 
		& + v_{\rm{blood}} C_{\rm{wb}}(t-t_d),
		\label{BFC}
	\end{aligned}
\end{equation}
where $t_d$ is the parameter representing the delayed time. With this model, all kinetic parameters $\tht=[v_{\rm{blood}}, K_1, k_2, k_3, t_d]^T$ are jointly estimated using the following nonlinear least-square time activity curve (TAC) fitting with a pre-determined $v_{\rm{bone}}$,
\beq
\hat{\tht} = \mathrm{argmim}_{\tht} \sum_{m=1}^M w_m \left(\tilde{C}_{\rm{PET}}(t_m) - C_{\rm{PET}}(t_m;\tht)\right)^2,
\eeq
where $\tilde{C}_{\rm{PET}}(t_m)$ is the measured tissue TAC extracted from dynamic PET mages. $M$ is the total number of dynamic frames. $t_m$ and $w_m$ are the time and the weight of the $m^{th}$ frame, respectively. A uniform weight (i.e. $w_m = 1$) is used in this study \cite{Wang2022a}. The classic Levenberg–Marquardt algorithm with 100 iterations is used to solve the optimization problem as the same used in \cite{Wang2012}.

To summarize, the standard tracer kinetic modeling approach ignores $v_{\rm{bone}}$ and its effect in Eq. \ref{BFC}, which may significantly overestimate the weight coefficient $(1-v_{\rm{bone}}-v_{\rm{blood}})$ if $v_{\rm{bone}}$ is significant, and consequently underestimate the $^{18}$F-FDG delivery rate $K_1$ by a factor of $\frac{1-v_{\rm{bone}}-v_{\rm{blood}}}{1-v_{\rm{blood}}}$, while $k_2$ and $k_3$ remain unchanged \cite{Holman2015}. An important macroparameter $K_i$ that includes $K_1$, representing the $^{18}$F-FDG net influx rate, is also included for the evaluation of kinetic quantification in this work,
\beq
K_i= \frac{K_1k_3}{k_2+k_3}.
\eeq
Due to the linear relationship between $K_1$ and $K_i$, the latter will similarly be underestimated by the same factor when $v_{\rm{bone}}$ is significant.
	\subsection{Quantitative Material Decomposition by DECT}
	One method to determine the bone fraction $v_{\rm{bone}}$ is to utilize X-ray DECT-based material decomposition. With high-energy CT $\muv^H$ and low-energy CT $\muv^L$, each image voxel $j$ can be decomposed as a set of material bases, such as air (A), soft tissue (S) or equivalently water, and bone (B):
	\beq
	\muv_j \triangleq \begin{pmatrix} \mu_j^H \\ \mu_j^L \end{pmatrix} = \U \ro_j,
	\eeq
	with
	\beq
	\U = \begin{pmatrix}
		\mu_A^H & \mu_S^H & \mu_B^H \\
		\mu_A^L & \mu_S^L & \mu_B^L
	\end{pmatrix}
	\quad \text{and} \quad
	\ro_j = \begin{pmatrix}
		\rho_{j,A} \\
		\rho_{j,S} \\
		\rho_{j,B}
	\end{pmatrix}.
	\eeq
	The matrix $\U$ consists of the linear attenuation coefficients of each basis material measured at both low and high energies. The coefficients $\rho_{j,k}$ with $k={A, S, B}$ represent the fraction of each basis material at the pixel $j$ and are constrained by $\sum_k \rho_{j,k} = 1$. $\ro_j$ can be estimated by using the following least-square optimization for each image voxel,
	\beq
	\hat{\ro}_j = \mathrm{argmim}_{\ro_j \geq 0} \|\muv_j - \U\ro_j\|^2.
	\label{MMD}
	\eeq
	
	Thus, the voxel-wise $v_{\rm{bone}}$ estimated from DECT (i.e., $\rho_{j,B}$) can be applied to Eq. \ref{SUV_BFC} and Eq. \ref{BFC} to correct bone marrow quantification in static and dynamic PET imaging. 
	
	\section{Evaluation Using Patient Datasets}
	\subsection{Study Subjects and Data Acquisition}
	\subsubsection{Total-body dynamic FDG-PET/CT} This study involved ten scans from five patients with metastatic genitourinary cancer (GUC) (3 males and 2 females; Age: 67±10 years) who underwent total-body dynamic scans following an injection of approximately 370 MBq of $^{18}$F-FDG using the uEXPLORER PET/CT system \cite{Cherry2017,Badawi2019,Spencer2021}. Prior Ethics Committee and Institution Review Board approval and written informed consent were obtained. Each patient underwent a baseline scan prior to their treatment (4 for anticancer immunotherapy and 1 for chemotherapy), followed by a follow-up scan approximately two weeks after starting treatment. All dynamic scans, lasting one hour each, were reconstructed into 29 frames of dynamic images (6×10 s, 2×30 s, 6×60 s, 5×120 s, 4×180 s, 6×300 s) using vendor-implemented time-of-flight ordered-subset expectation maximization algorithm with 4 iterations and 20 subsets. Corrections for randoms, scatter, attenuation, deadtime, decay, and normalization, were all applied. The image dimension for each frame was 150×150×486 with a voxel size of 4×4×4 mm$^3$ with no post-reconstruction image smoothing. A whole-body X-ray CT scan at a tube voltage of 140 kVp ($\sim$50 mAs with automatic dose modulation) was performed for the purposes of attenuation correction and localization.
	
	\subsubsection{DECT and Material Decomposition} In addition to the CT scan at 140 kVp, each patient underwent the other CT scan at 80 kVp ($\sim$100 mAs with automatic dose modulation), allowing for the formation of DECT pairs. The two CT scans were conducted consecutively, with the second scan taking place immediately after the first.  We applied a deep-learning-based algorithm \cite{Li2020} to improve the image quality of the 80 kVp image, followed by a B-spine registration algorithm \cite{Wang2015} to align the two CT images. Note that, the average bias of Hounsfield Unit (HU) quantification in bone marrow regions with and without denoising is less than 2\%. Material decomposition was performed on DECT images to determine the air, soft-tissue, and bone fractions for each voxel. We set the material density values for air, soft-tissue, and bone as 0.0012 g/cm$^3$, 1.06 g/cm$^3$, and 2.4 g/cm$^3$, respectively, to pre-determine the material basis matrix $\U$ (Eq. \ref{MMD}). 
	
	\subsubsection{Extraction of bone marrow and blood time activity curves (TAC)} 
\begin{figure}[t]
		\centering
		\subfloat[]{\includegraphics[trim=0cm 0cm 0cm 0cm, width=1.7in]{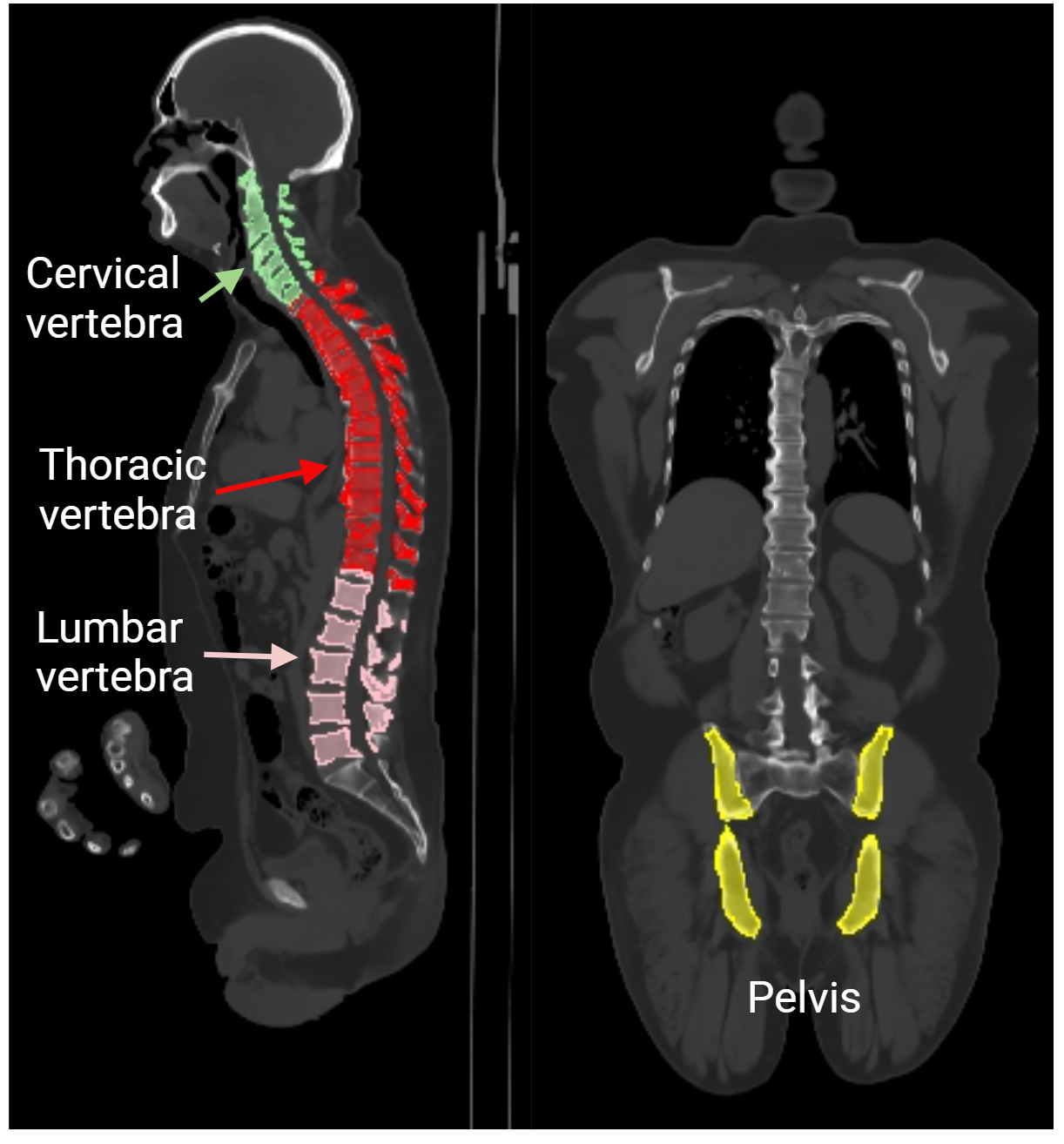}}
		\subfloat[]{\includegraphics[trim=0cm 0cm 0cm 0cm, width=1.7in]{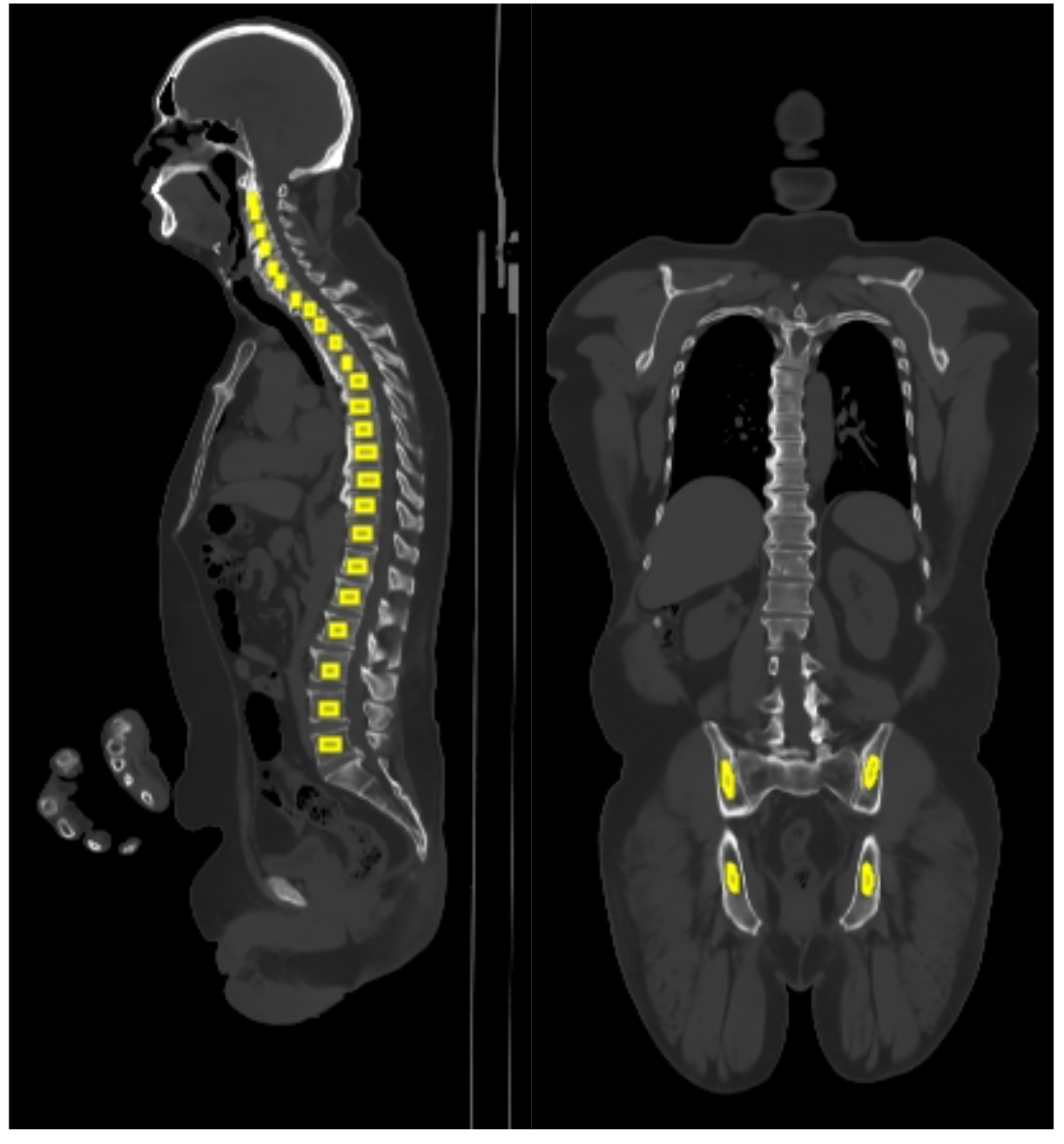}}
		\caption{Illustration of bone marrow ROI placement. (a) Automatic segmentation of spine and pelvis regions enabled by TotalSegmentator. (b) Manual ROI delineation based on (a). CT HU window: [-300, 1200]}
		\label{total}	
	\end{figure}

The cervical vertebra (CV: C$_1$-C$_7$), thoracic vertebra (TV: T$_1$-T$_{12}$), lumbar vertebra (LV: L$_1$-L$_5$), and pelvis regions were first segmented using the TotalSegmentor toolbox \cite{Wasserthal2023} installed in 3D Slicer \cite{Fedorov2012}, as shown in Fig. \ref{total}a. Fig. \ref{total}b further presents the manual delineation for all regions of interest (ROIs) based on the auto-segmentation results by aligning the CT, dynamic PET, and last 5-minute static PET images. The four averaged regions comprised CV from seven ROIs, TV from twelve ROIs, LV from five ROIs, and pelvis from four ROIs, which were used to extract the global bone marrow TACs from the dynamic images with lower noise. An additional ROI was placed in the descending aorta region to extract the image-derived input function by considering early and late frames simultaneously. The SUV was calculated using the data of last 5 minutes (i.e. 55-60 min).
	
		\begin{figure}[t]
		\vspace{-0pt}
		\footnotesize
		\centering
		{\includegraphics[trim=0cm 0cm 0cm 1cm, clip,width=3.4in]{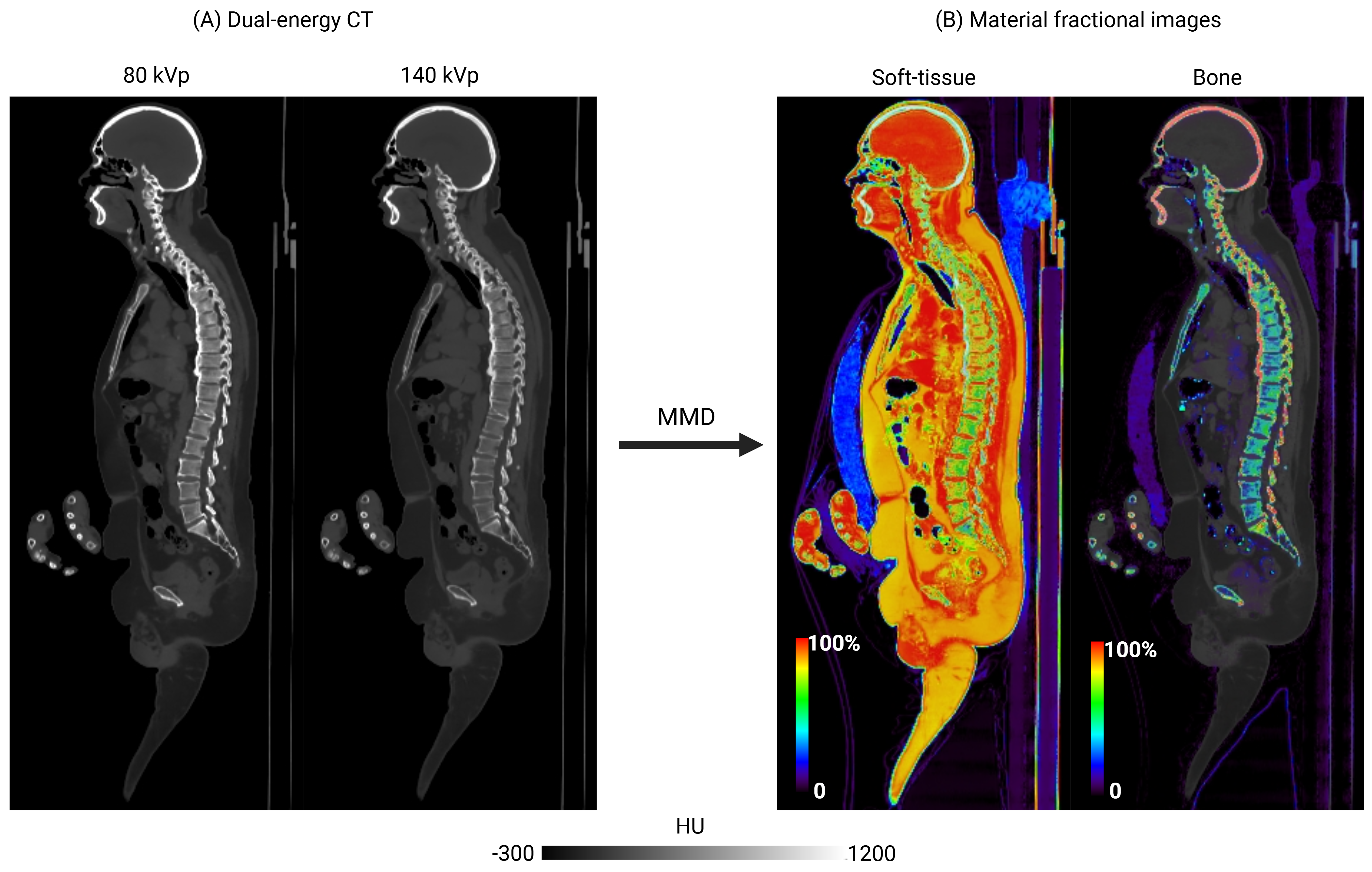}}
		\caption{Visualization of X-ray DECT images (left, shown in HU window [-300, 1200]) and corresponding soft-tissue and bone fractional images overlaid on the 140 kVp CT image (right).}
		\label{DECT and MMD}
		\vspace{-0pt}
	\end{figure} 
	\subsection{Statistic Analysis}
	\subsubsection{Investigation of Bone Fraction in Bone Marrow} We first reported the bone fraction (mean ± standard deviation (SD)) in each bone marrow region across all scans (a total of ten scans: five patients with both baseline and follow-up scans). A paired t-test was used to evaluate the statistical differences in bone fractions (1) among various regions, and (2) between baseline and follow-up scans within each region. We then conducted a linear correlation analysis between the bone fraction of each region and patient characteristics, e.g., age, body mass index (BMI), and glucose level (Glu) on five baseline scans. The difference in bone fractions between males and females was also compared using the unpaired t-test. 
	\subsubsection{Impact on Metabolic Quantification of Bone Marrow} To evaluate the effect of BFC on the quantification of bone marrow using static PET, a group comparison was first performed for SUVs of different spine and pelvis bone marrows using the paired t-test. We then performed the same comparisons for $^{18}$F-FDG delivery rate $K_1$ and net influx rate $K_i$ to investigate the bone fraction effect on the kinetic quantification of bone marrow. Here, all ten scans from 5 patients were used. While each scan was analyzed independently, we recognized that scan pairs from the same patient are not fully independent. Given that the analysis in this study focused on the correction effect rather than longitudinal changes, this assumption was considered appropriate. P values of less than 0.05 were considered statistically significant.
	\subsubsection{Parametric Imaging of Bone Marrow} In addition to ROI-based analysis, parametric images of SUV, $K_1$, and $K_i$ from one patient without and with BFC, as well as their difference images in percentage, were generated for demonstration. Kernel smoothing, as described in \cite{Wang2015a,Wang2022a}, was applied to the generation of $K_1$ and $K_i$ images. A bone mask was employed to visualize only the regions of bone marrow. 
	
	\section{Results}
	\subsection{DECT and Material Fractional Images}
	
	An example of DECT image pairs, one at 80 kVp and the other at 140 kVp, is shown in the left of Fig. \ref{DECT and MMD}. Their corresponding soft-tissue and bone fractional images are further shown in the right of Fig. \ref{DECT and MMD}. In the soft-tissue fractional image, the brain and liver regions presented almost 100\% soft-tissue component, while the muscle region (orange) appeared slightly lower. This phenomenon is expected because the muscle contains both soft tissue and fat. From the bone fractional image, the spine bone marrow region indicated a moderate bone fraction (average: $\sim$15\%), while the skull exhibited a higher bone fraction (average: $\sim$70\%), which matches the physiological knowledge. 
	
		\begin{figure}[t]
		\vspace{-0pt}
		\footnotesize
		\centering
		\subfloat[]{\includegraphics[trim=0cm 0cm 8.3cm 2cm, clip, width=2.5in]{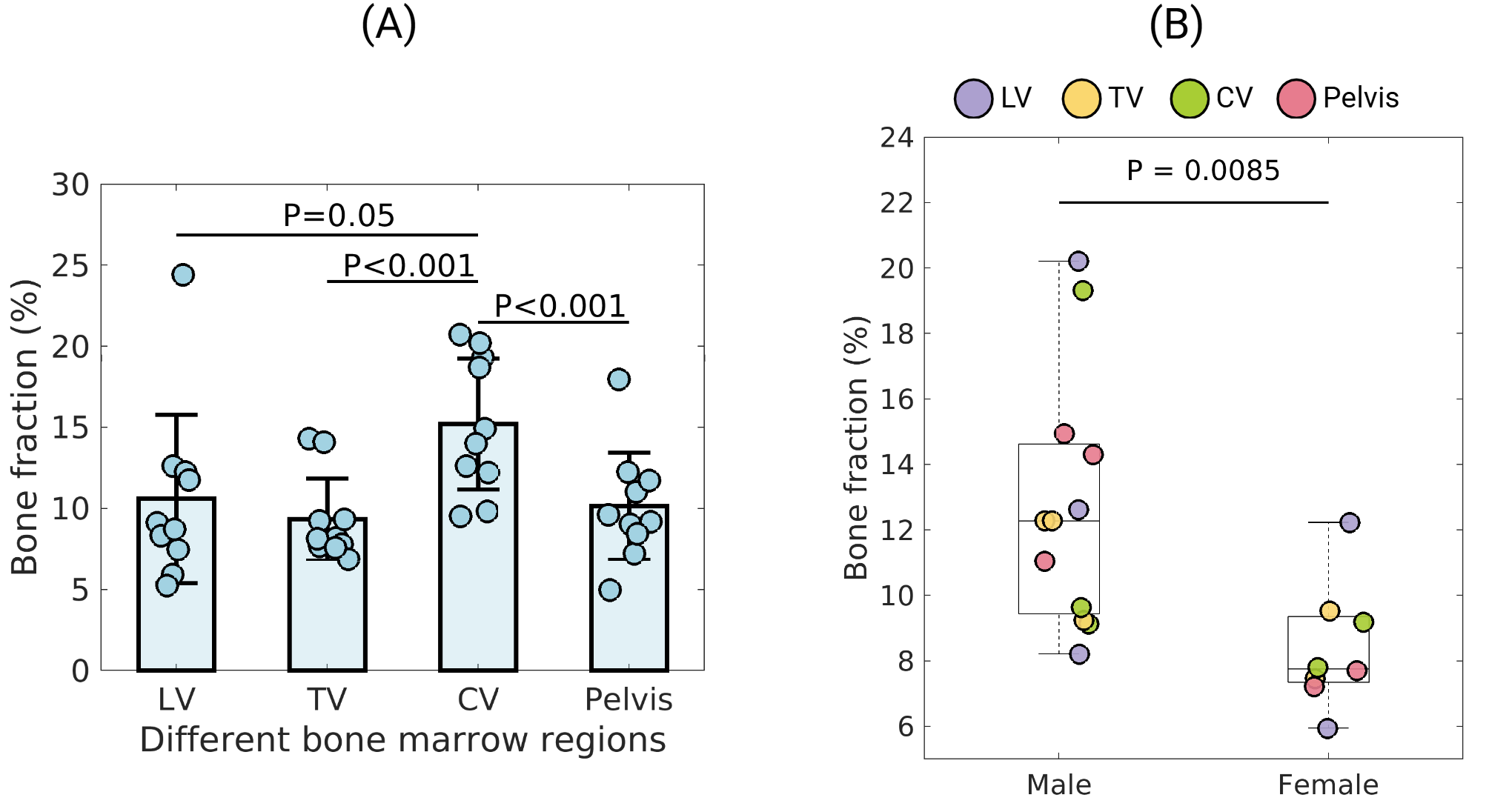}}\\
		\subfloat[]{\includegraphics[trim=9.6cm 0cm 0cm 0.6cm, clip, width=2.2in]{bone_fraction_new.png}}
		\caption{Results of bone fractions for different regions of interest: cervical vertebra (CV), thoracic vertebra (TV), lumbar vertebra (LV), and pelvis. (a) Bar plots of bone fraction in different regions across all the ten scans. Only P values of paired t-test between CV and other regions were displayed. The other P values were all larger than 0.2. (b) Boxplot comparison of bone fraction between males and females using the unpaired t-test.}
		\label{BF}
		\vspace{-0pt}
	\end{figure}
	\subsection{Knowledge of Bone Fraction in Bone Marrow}
	Fig. \ref{BF}a illustrates the differences in bone fraction across various bone marrow regions from all the ten scans. On average, the bone fraction in these regions was approximately 11\%, but up to 25\%. The mean ± SD of bone fractions for LV, TV, CV, and pelvis were 10.6\% ± 5.2\%, 9.3\% ± 2.4\%, 15.2\% ± 5\%, and 10.2\% ± 3.3\%, respectively. The bone fraction in CV was statistically higher than in TV and pelvis (P$ < $0.001) and was marginally higher than in LV (P=0.05). We found no significant difference in bone fraction between baseline and follow-up scans for each bone marrow region, which is not surprising because the average interval between the two scans is two weeks.
	
	Fig. \ref{BF}b shows an unpaired group comparison of bone fraction in each bone marrow region stratified by sex. This analysis revealed a statistically significant higher bone fraction in males as compared to females (P$ < $0.01). Our findings might align with the existing research that suggests sex hormones contribute to decreased bone fraction in females \cite{Zhang2024}.
	
The linear regression analysis results examining the relationship between bone fractions in different bone marrow regions (LV, TV, CV, and pelvis) and patient characteristics (BMI, and Glu) did not show any statistically significant correlations (P$>$0.2). However, there was a trend indicating a decrease in bone fraction with age (R=0.85, P=0.06; figure not shown).

		\begin{figure*}[h]
		\vspace{-0pt}
		\footnotesize
		\centering
		{\includegraphics[trim=0cm 0cm 0cm 0cm, clip,width=5in]{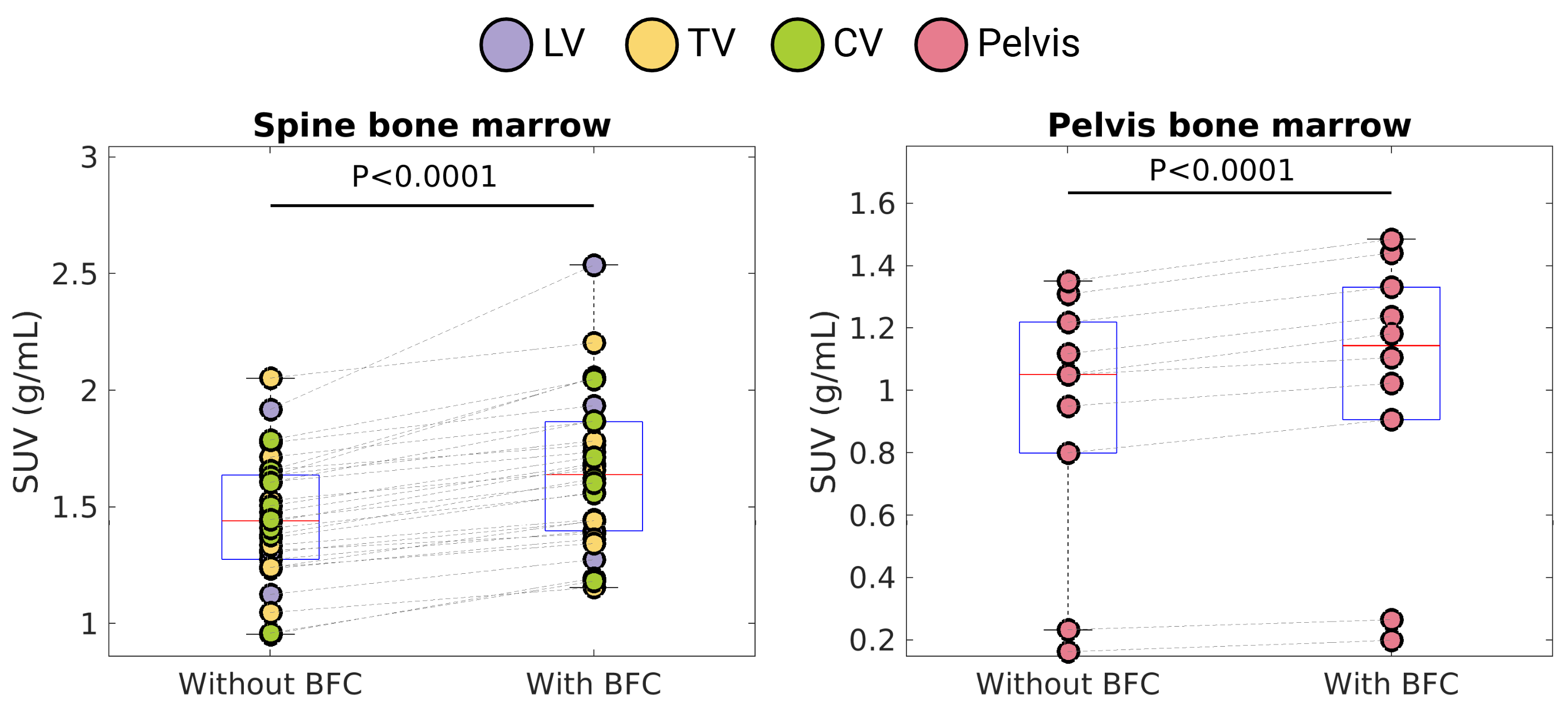}}
		\caption{Comparison between SUV without BFC and SUV with BFC for each spine and pelvis bone-marrow region across ten scans. The paired lines and P values of paired t-test were included.}
		\label{SUV}
		\vspace{-0pt}
	\end{figure*} 
	
	\subsection{SUV Quantification of Bone Marrow}
	The group comparisons of SUV values estimated without and with BFC for spine (LV, TV, CV) and pelvis bone marrow ROIs across all the ten scans are shown in Fig. \ref{SUV}. The increase of SUV by BFC was 12.3\%±7.2\%, 10.4\%±3.2\%, 18.2\%±5.6\%, and 11.5\%±4.3\% for LV, TV, CV, and pelvis, respectively. These results indicated a statistically significant effect of BFC on the SUV calculations (P$<$0.0001).

	\subsection{Kinetic Quantification of Bone Marrow}
\begin{figure}[t]
		\vspace{-0pt}
		\footnotesize
		\centering
		\subfloat[]{\includegraphics[trim=0cm 0cm 7.5cm 0cm, clip, width=2.5in]{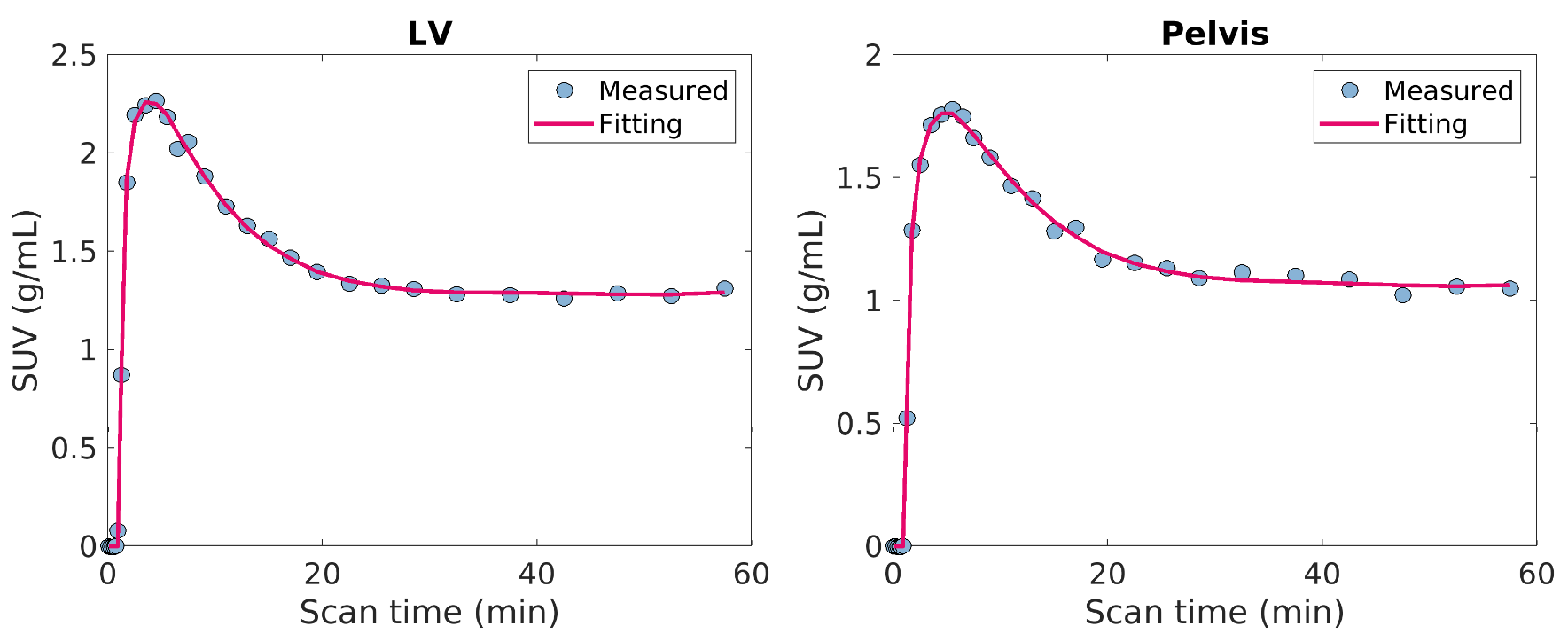}}\\
		\subfloat[]{\includegraphics[trim=7.5cm 0cm 0cm 0cm, clip, width=2.5in]{K_fitting.png}}
		\caption{Examples of $^{18}$F-FDG TAC fitting for (a) a LV bone-marrow region and (b) a pelvis bone-marrow region using the 2-tissue irreversible model with considering the bone fraction effect.}
		\label{K_fitting}
		\vspace{-0pt}
	\end{figure}

	Fig. \ref{K_fitting} first shows two examples of $^{18}$F-FDG TAC fitting for a spine (e.g., LV as an example) and a pelvis ROI using the 2-tissue irreversible compartmental model with considering the bone fraction effect. The fitted curves demonstrated a good agreement with the measured data. ROI fitting on other subjects also showed similar behavior. Our results were consistent with the previous finding \cite{Wang2023}, suggesting that the 2-tissue irreversible model is appropriate for the kinetic fitting of bone marrow. Note that, the fitting behavior without and with BFC was the same because the effect of including the bone fraction $v_{\rm{bone}}$ only caused a scaling on $K_1$ without changing the TAC fitting.
	
\begin{figure*}[t]
		\vspace{-0pt}
		\footnotesize
		\centering
		{\includegraphics[trim=0cm 0cm 0cm 0cm, clip,width=5in]{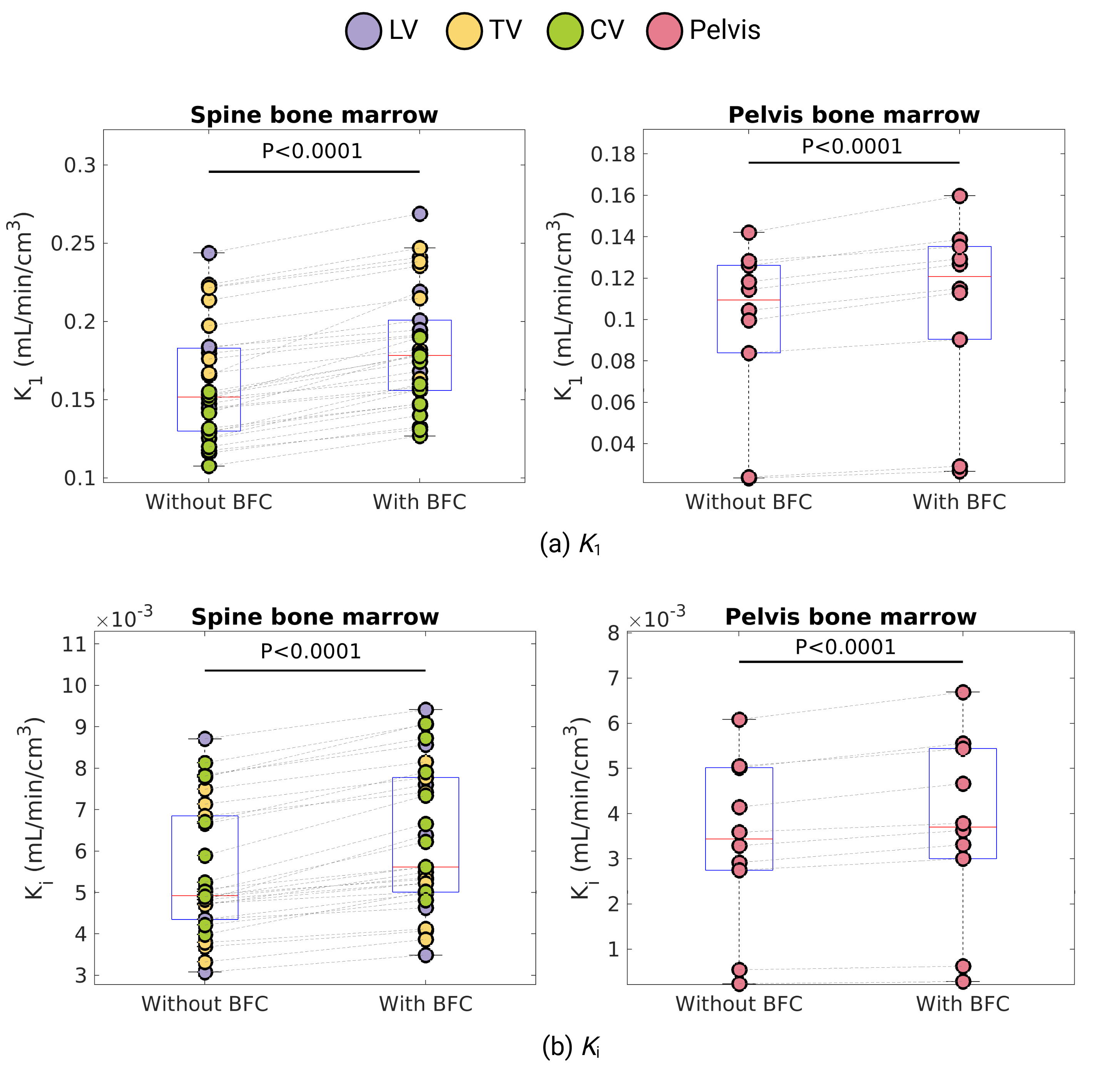}}
		\caption{Effect of BFC on (a) FDG $K_1$ and (b) $K_i$ quantification for spine and pelvis marrow regions across all the ten scans. The paired lines and P values of paired t-test were included.}
		\label{K}
		\vspace{-0pt}
	\end{figure*}

	The comparative analysis of $K_1$ and $K_i$, both in the absence and presence of BFC, across various bone marrow ROIs is illustrated in Fig. \ref{K}. Consistent with the observation in SUV, the changes in them were also found to be statistically significant (all P$<$0.0001). The increase in $K_1$ and $K_i$ attributable to BFC were 12.4\% ± 7.2\% in LV, 10.4\% ± 3.2\% in TV, 18.8\% ± 5.7\% in CV, and 11.5\% ± 4.2\% in pelvis, respectively. The small differences in increases between these two kinetic parameters and SUV are due to the consideration of $v_{\rm{blood}}$, which was approximately 1\% in LV, 0.5\% in TV, 2\% in CV, and 0.5\% in pelvis, respectively.
	
	\subsection{Demonstration of Bone Marrow Multiparametric Imaging}
\begin{figure}[t]
	\vspace{-0pt}
	\footnotesize
	\centering
	{\includegraphics[trim=0cm 0cm 0cm 0cm, clip,width=3.4in]{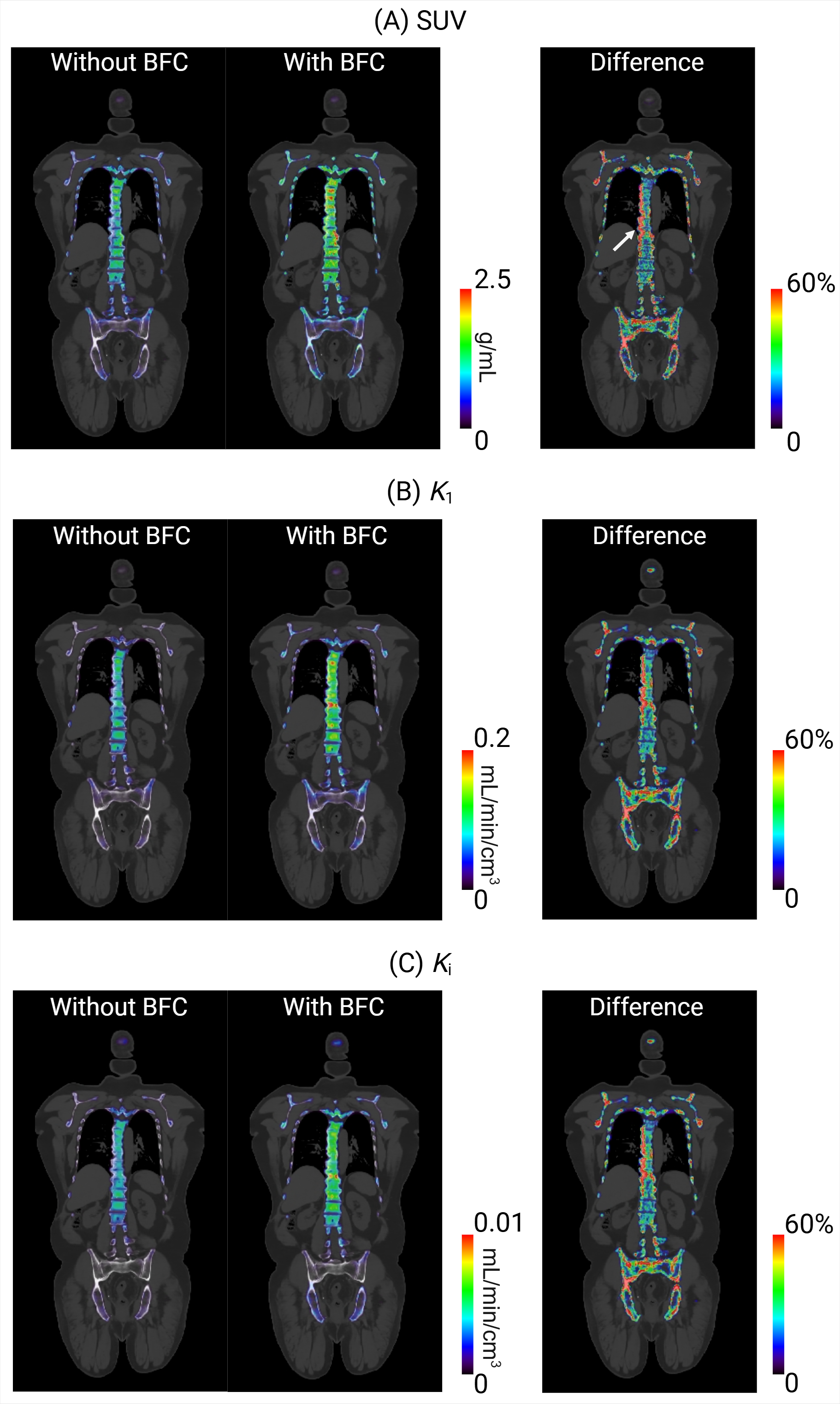}}
	\caption{Total-body bone marrow images of (a) SUV, (b) $K_1$, and (c) $K_i$ generated without BFC and with BFC from one representative patient. The corresponding difference images in percentage were included. All the images were overlaid on the 140 kVp image (shown in HU window [-300, 1200]).}
	\label{multiP}
	\vspace{-0pt}
	\end{figure}

Total-body parametric images of SUV, $K_1$ and $K_i$ of bone marrow generated without BFC and with BFC are demonstrated in Fig. \ref{multiP} for one representative patient. The corresponding percentage difference images are also included. All these images are overlaid on the corresponding 140 kVp CT image. As expected from the ROI-based analysis, all three parametric images without BFC demonstrated an underestimation in bone marrow regions as compared to that with BFC. The median value in the difference image was around 21\%. The large difference regions (white arrow) correspond to the cortical bone regions, which are expected to have a relatively higher bone fraction. Those regions were excluded from our ROI analysis, as the actual bone marrow is primarily located within the spongy bone region.

	\section{Discussions}

In this study, we demonstrated a bone-led tissue composition effect for bone marrow imaging and then developed a correction approach that integrates the bone volume fraction estimated from DECT material decomposition into the PET measurements (Eq. \ref{SUV_BFC} and Eq. \ref{BFC}). We evaluated the impact of this correction method on metabolic quantification of bone marrow in static and dynamic imaging using total-body dynamic $^{18}$F-FDG PET scans. To our knowledge, this was the first study to introduce bone fraction correction into metabolic PET quantification by using DECT. 

The results of bone fraction estimation, shown in Fig. \ref{BF}a, indicated that the CV region had a higher bone fraction compared to other regions. This is consistent with the higher bone density reported in the cervical spine, which is often attributed to biomechanical demands \cite{Simion2023,Anderst2011}. We observed a negative correlation between bone fraction and age (R=0.85, P=0.06), which is consistent with findings reported in the literature \cite{Chen2013}. The further comparison stratified by sex (Fig. \ref{BF}b) also aligned with the existing knowledge: generally, the bone density in females is lower than in males \cite{Zhang2024}. All these results may suggest that the material decomposition approach used in this work is reasonable and feasible.

The ROI-based group comparisons demonstrated that the  SUV, $K_1$, and $K_i$ values in different bone marrow regions were all significantly underestimated without BFC (Figs. \ref{SUV} and \ref{K}). The larger the bone fraction, the more significant the difference between them observed. These results indicate the importance of BFC; otherwise, there would be a significant error in the metabolic quantification of bone marrow. The parametric image visualization (Fig. \ref{multiP}) further confirmed the similar result. Of note, only the $^{18}$F-FDG delivery rate $K_1$ and any other macroparameters of which the calculation involves $K_1$ (e.g., $K_i$), need to be corrected, as $k_2$ and $k_3$ remain unchanged with and without BFC.

The proposed BFC method shares a concept similar to the tissue fraction correction used in lung PET imaging  \cite{Holman2015}. The latter utilized the single X-ray CT image to determine air fraction. For our bone marrow imaging applications, the X-ray DECT method is necessary to decompose the fraction of bone from soft-tissue, which distinguishes our work from earlier studies \cite{Holman2015,Lambrou2011}. 

	While this work used X-ray DECT to demonstrate the necessity of BFC for bone marrow quantification, the broad use of X-ray DECT may be limited with current PET/CT. The integration of X-ray DECT with PET/CT is not trivial due to the need for an additional X-ray CT scan, which may either increase the radiation dose or require a hardware upgrade from standard single-energy X-ray CT. The PET-enabled DECT method can address these issues by combining a high-energy 511 keV $\gamma$-ray CT (gCT) reconstructed from PET emission data with the already-existing X-ray CT to form dual-energy imaging \cite{Wang2020}. With advanced reconstruction and scatter correction methods, the quality of the gCT image has been substantially improved, which in turn benefits the material decomposition as demonstrated through simulation data\cite{Li2022}, physical phantom data \cite{Li2024,Zhu2023}, and real patient scans \cite{Zhu2025}. This warrants a future study to evaluate BFC for bone marrow quantification using the PET-enabled DECT method, which requires only a single CT scan.

	We observed some limitations in this study. First, the sample size of subjects scanned with both $^{18}$F-FDG PET and DECT was small in this pilot study. As the first demonstration, this work mainly focused on the technical development and demonstration of a method for correcting the metabolic quantification of bone marrow. Future research will involve a larger dataset to further improve and refine the correction method, for example, including correction for the red/yellow marrow fraction. Second, the current investigation focused on the spine and pelvis marrow, where defining the ROI is more straightforward. In future studies, we plan to include other bone marrow regions, such as the sternum and the spongy ends of long bones (e.g., the femur and humerus). Third, the estimation of bone fraction may be affected by several factors, including poor CT image quality and large misalignment between PET and CT. Although these issues did not happen in our pilot study, it is worth developing advanced technical methods to address them when including more datasets in the future. Finally, we noted that the effect of bone fraction could lead to significant differences in metabolic quantification in PET imaging. However, it remains unclear whether the corrected bone marrow quantification will have an impact on clinical applications, such as in detecting bone marrow involvement during cancer staging or in assessing responses to anti-cancer immunotherapy. Therefore, in future work, we plan to recruit more cancer patients who will undergo baseline PET, follow-up PET, and dual-energy CT scans to conduct further clinical studies. Both static and dynamic PET measures of bone marrow, with and without bone fraction correction, will be analyzed with the clinical outcomes.

	\section{Conclusion}
	In this work, we have developed a bone fraction correction method to evaluate the effect of bone fraction on metabolic $^{18}$F-FDG-PET quantification of bone marrow using dual-energy CT material decomposition. Our results suggest that current SUV and kinetic quantification of FDG delivery rate $K_1$ and net influx rate $K_i$ in bone marrow are likely to be significantly underestimated in PET due to the substantial presence of bone within a voxel or region of interest. This underscores the critical importance of implementing bone fraction correction for metabolic quantification of bone marrow, which may have wide applications in blood cancer staging and anti-cancer immunotherapeutic response assessment.  
	\section*{Acknowledgment}

	We acknowledge contributions of team members in the EXPLORER Molecular Imaging Center, UC Davis.

\end{document}